\documentclass[fleqn,usenatbib]{mnras}

\usepackage{newtxtext,newtxmath}
\usepackage{adjustbox}

\usepackage[T1]{fontenc}

\DeclareRobustCommand{\VAN}[3]{#2}
\let\VANthebibliography\thebibliography
\def\thebibliography{\DeclareRobustCommand{\VAN}[3]{##3}\VANthebibliography}

\usepackage{graphicx}	
\usepackage{amsmath}	
\usepackage{subfig}
\usepackage{booktabs}
\usepackage{multirow}

\title[Redshift Estimation]{Estimation of redshift and associated uncertainty of Fermi/LAT extra-galactic sources with Deep Learning}

\author[Gharat et al.]{
Sarvesh Gharat,$^{1}$\thanks{E-mail: sarveshgharat19@gmail.com}
Abhimanyu Borthakur,$^{2}$
and Gopal Bhatta$^{3}$
\\
$^{1}$Centre for Machine Intelligence and Data Science, Indian Institute of Technology Bombay, 400076, Mumbai, India\\
$^{2}$Department of Electronics and Communication Engineering, Manipal Institute of Technology, 576104, Karnataka, India\\
$^{3}$ Janusz Gil Institute of Astronomy, University of Zielona Góra, ul. Szafrana 2, 65-516 Zielona Góra, Poland
}

\date{Accepted XXX. Received YYY; in original form ZZZ}

\pubyear{2023}

\begin{document}
\label{firstpage}
\pagerange{\pageref{firstpage}--\pageref{lastpage}}
\maketitle

\begin{abstract}
With the advancement of technology, machine learning-based analytical methods have pervaded nearly every discipline in modern studies. Particularly, a number of methods have been employed to estimate the redshift of gamma-ray loud active galactic nuclei (AGN), which are a class of supermassive black hole systems known for their intense multi-wavelength emissions and violent variability. Determining the redshifts of AGNs is essential for understanding their distances, which, in turn, sheds light on our current understanding of the structure of the nearby universe. However, the task involves a number of challenges such as the need for meticulous follow-up observations across multiple wavelengths and astronomical facilities. In this study, we employ a simple yet effective deep learning model with a single hidden layer having $64$ neurons and a dropout of $0.25$ in the hidden layer,  on a sample of AGNs with known redshifts from the latest AGN catalog, 4LAC-DR3, obtained from Fermi-LAT. We utilized their spectral, spatial, and temporal properties to robustly predict the redshifts of AGNs as well quantify their associated uncertainties, by modifying the model using two different variational inference methods. We achieve a correlation coefficient of $0.784$ on the test set from the frequentist model and $0.777$ and $0.778$ from both the variants of variational inference, and, when used to make predictions on the samples with unknown redshifts, we achieve mean predictions of $0.421, 0.415$ and $0.393$, with standard deviations of $0.258$, $0.246$ and $0.207$ from the models, respectively.



\end{abstract}



\begin{figure*}
    \centering
    \includegraphics[scale = 0.85]{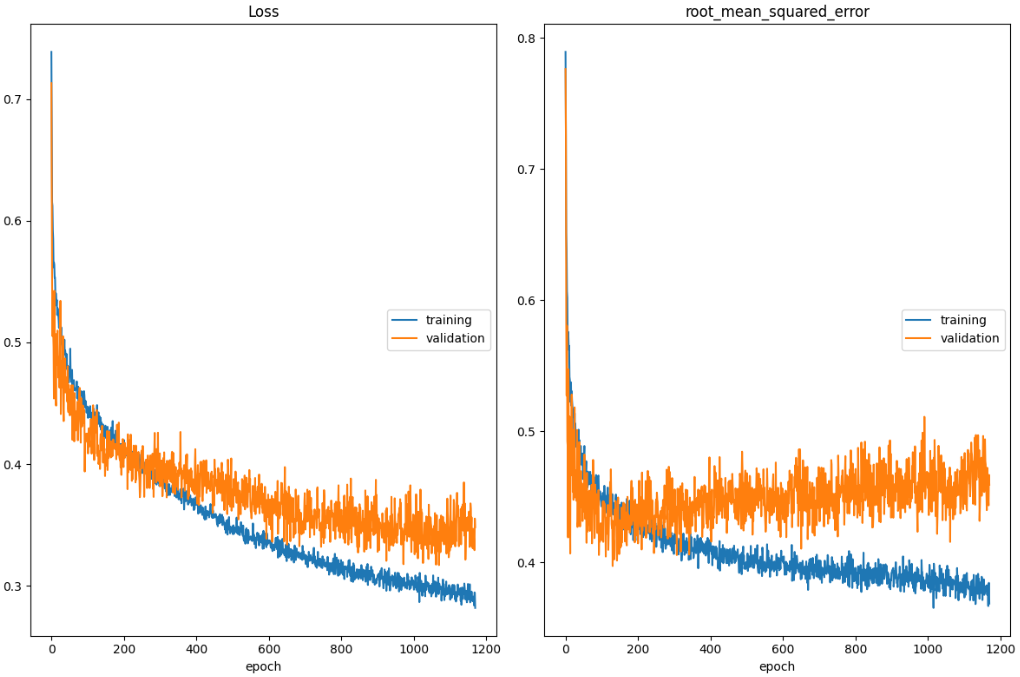}
    \caption{Plots for Epochs vs Loss (MAE) and RMSE for Variational Inference (Flipout Estimator)}
    \label{fig:flipout}
\end{figure*}

\begin{figure*}
    \centering
    \includegraphics[scale = 0.85]{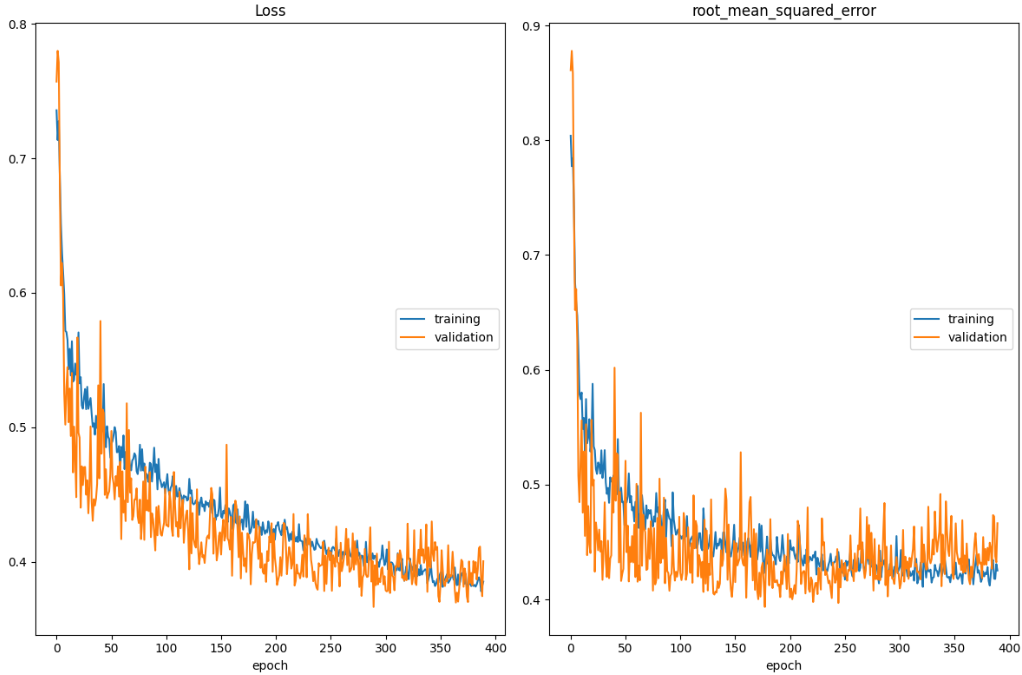}
    \caption{Plots for Epochs vs Loss (MAE) and RMSE for Variational Inference (Reparameterization Estimator)}
    \label{fig:reparameterization}
\end{figure*}

\begin{figure*}
    \centering
    \includegraphics[scale=0.50]{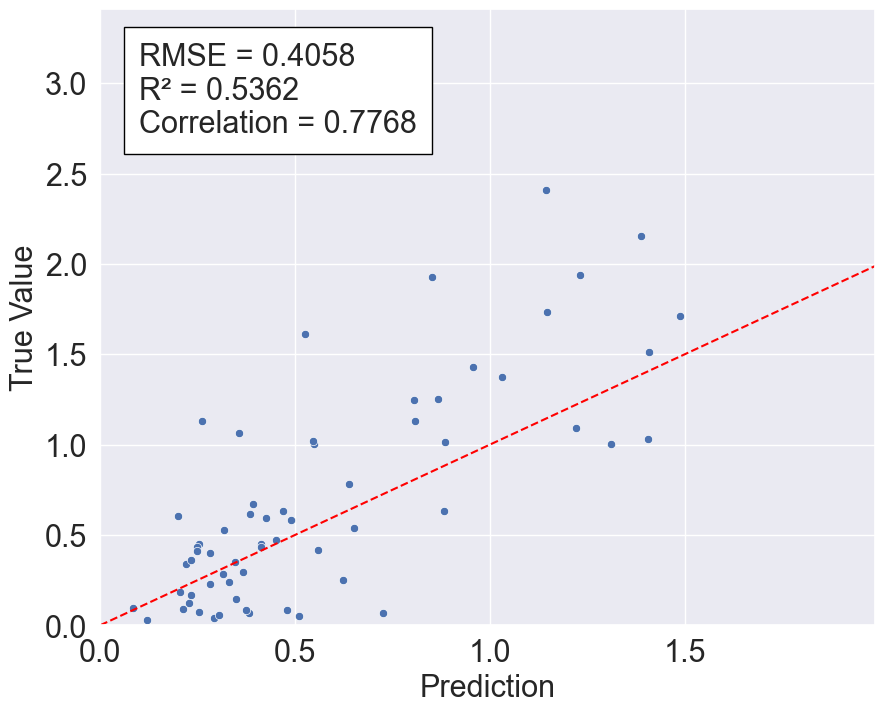}
    \caption{Scatter relation between the true value and the predicted mean value using Variational Inference (Flipout Estimator). The red diagonal represents a perfect prediction.}
    \label{fig:flipout_scatter}
\end{figure*}

\begin{figure*}
    \centering
    \includegraphics[scale=0.50]{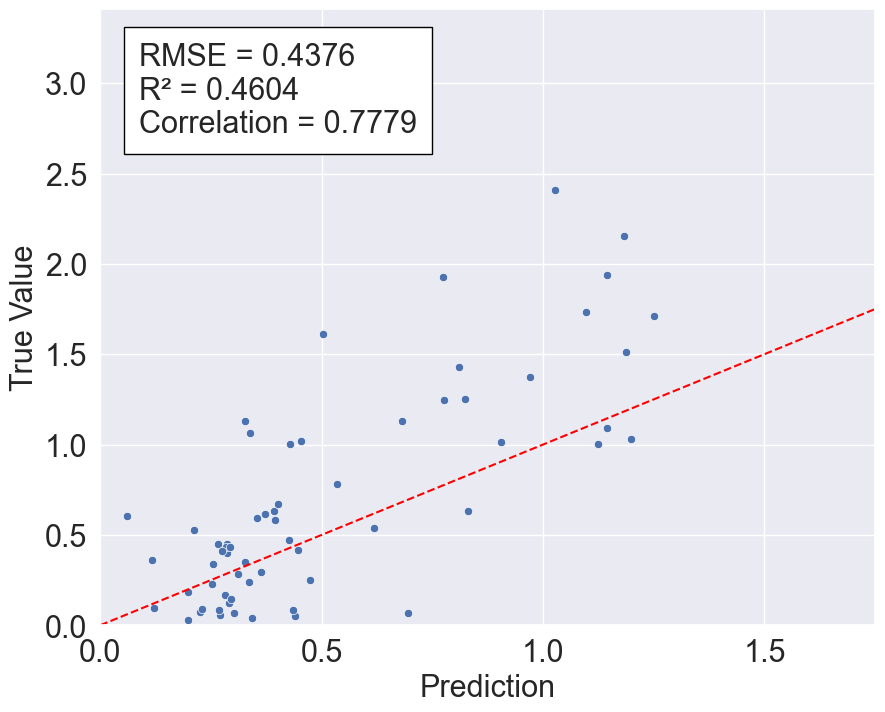}
    \caption{Scatter relation between the true value and the predicted mean value using Variational Inference (Reparameterization Estimator). The red diagonal represents a perfect prediction.}
    \label{fig:reparameterized_scatter}
\end{figure*}

\begin{figure*}
    \centering
    \includegraphics[scale=0.50]{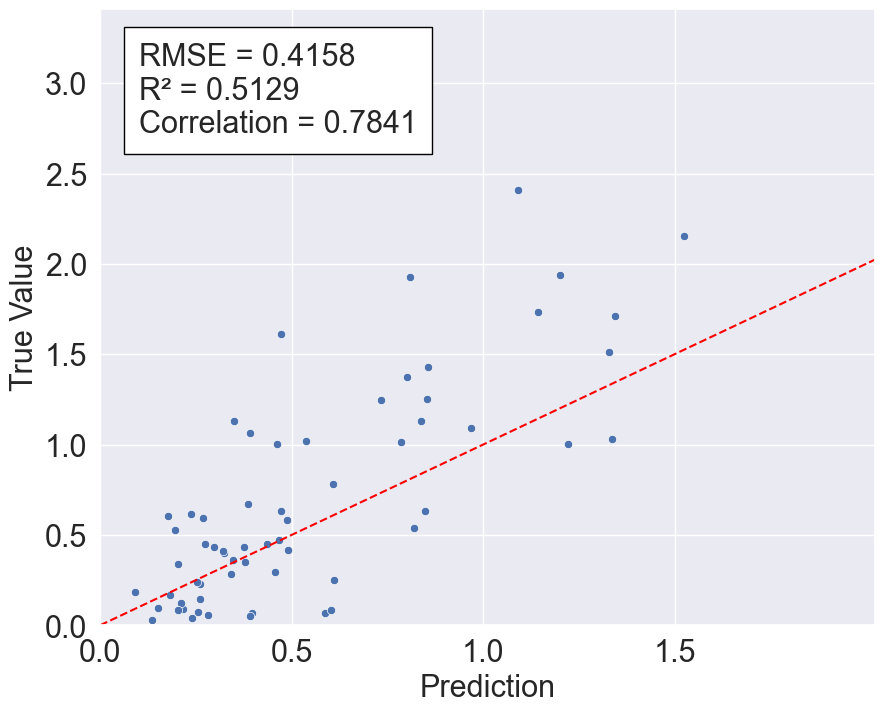}
    \caption{Scatter relation between the true value and the predicted value using Frequentist model. The red diagonal represents a perfect prediction.}
    \label{fig:frequentist_scatter}
\end{figure*}

\begin{figure*}
    \centering
    \includegraphics[scale=0.325]{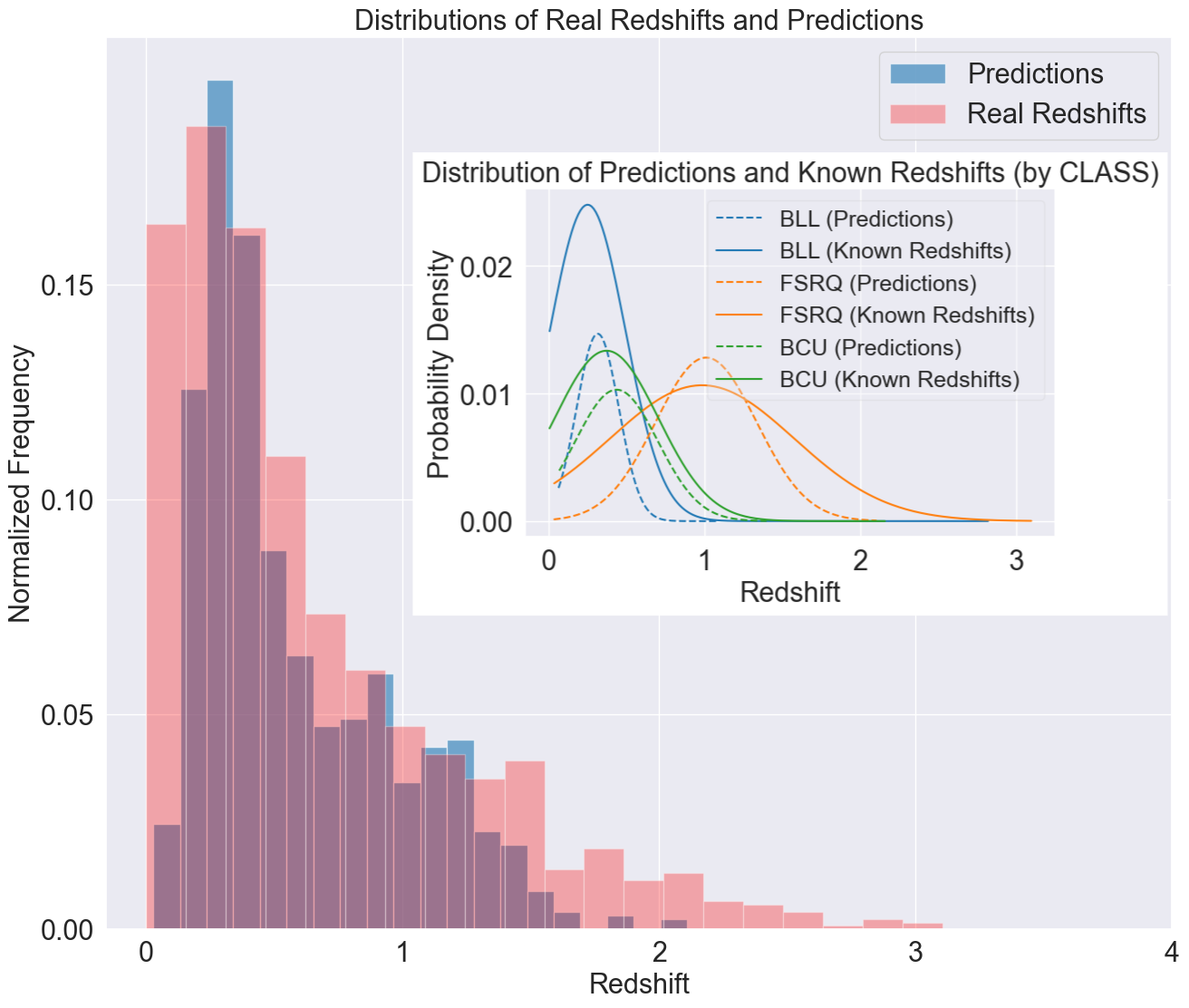}
    \caption{Variational Inference (Flipout Estimator) - Comparison between Predicted Mean Redshift and True Redshift using Histograms. The distribution of the redshift values for both the known and predicted redshifts, disaggregated by the "CLASS" feature, is shown. Here, only those classes with more than 50 samples are represented.}
    \label{fig:Flipout_redshift}
\end{figure*}

\begin{figure*}
    \centering
    \includegraphics[scale=0.325]{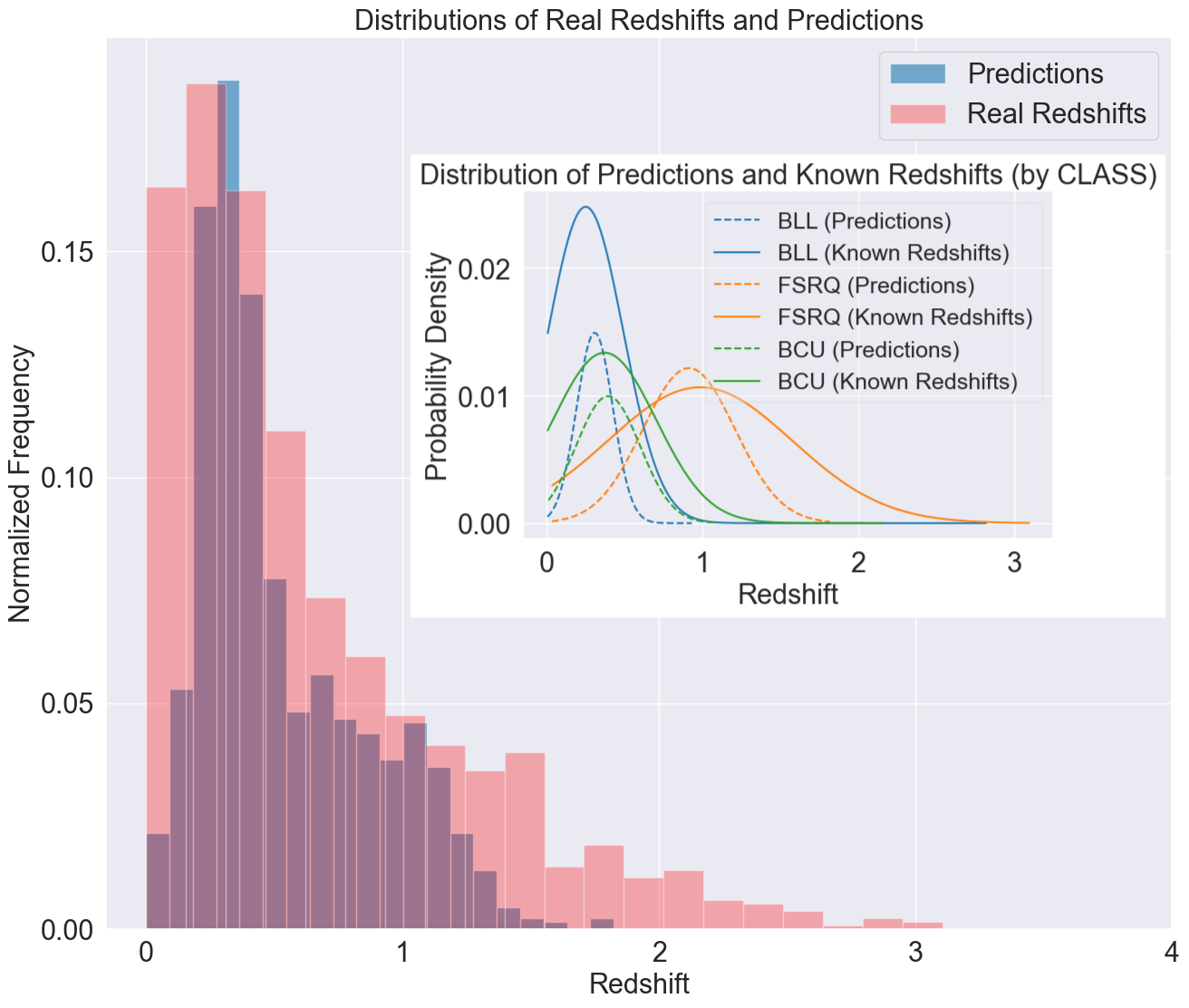}
    \caption{Variational Inference (Reparameterized Estimator) - Comparison between Predicted Mean Redshift and True Redshift using Histograms. The distribution of the redshift values for both the known and predicted redshifts, disaggregated by the "CLASS" feature, is shown. Here, only those classes with more than 50 samples are represented.}
    \label{fig:Reparameterized_redshift}
\end{figure*}

\begin{figure*}
    \centering
    \includegraphics[scale=0.325]{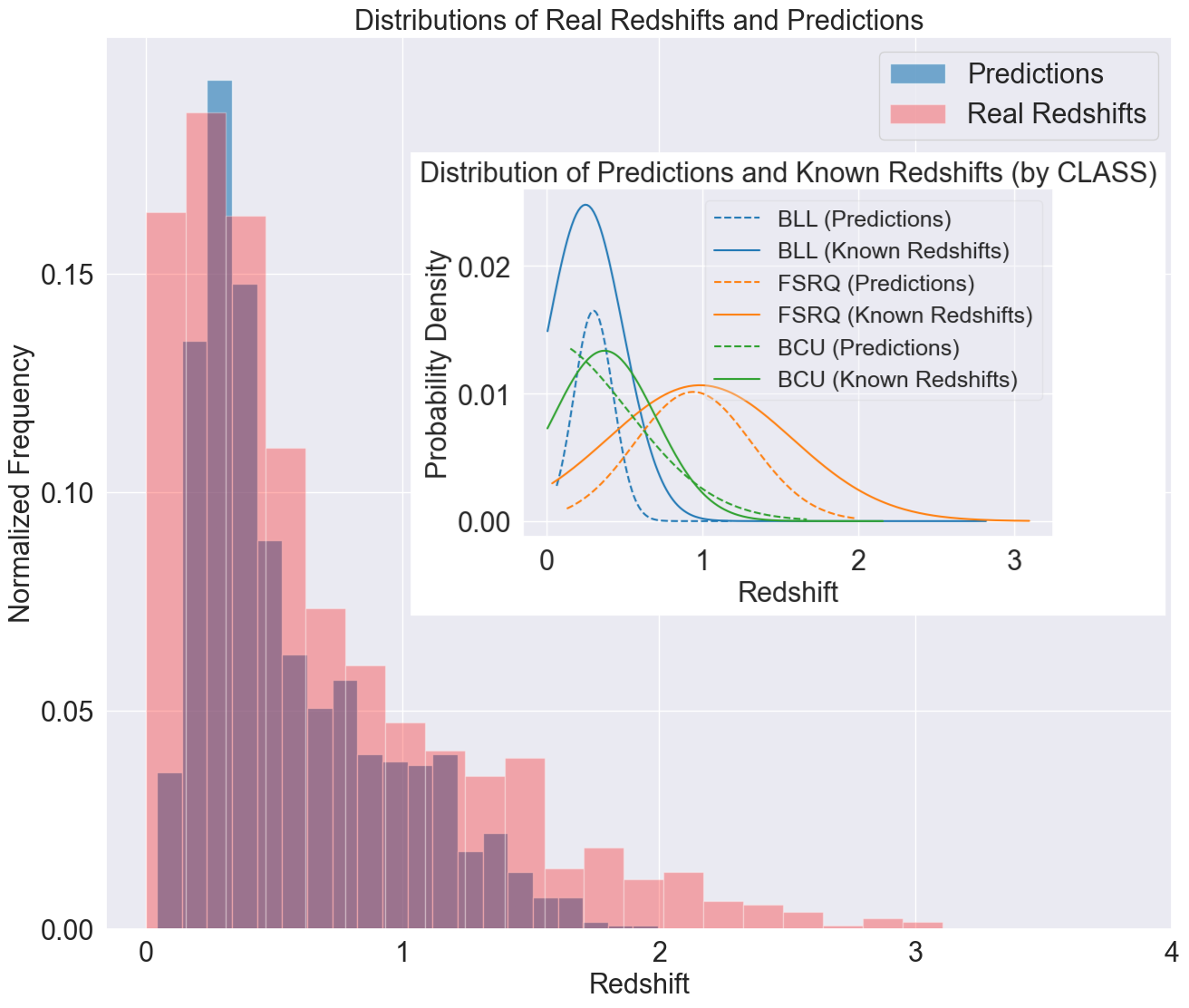}
    \caption{Frequentist model - Comparison between Predicted Mean Redshift and True Redshift using Histograms. The distribution of the redshift values for both the known and predicted redshifts, disaggregated by the "CLASS" feature, is shown. Here, only those classes with more than 50 samples are represented.}
    \label{fig:Frequentist_redshift}
\end{figure*}

\begin{figure*}
    \centering
    \includegraphics[scale=0.325]{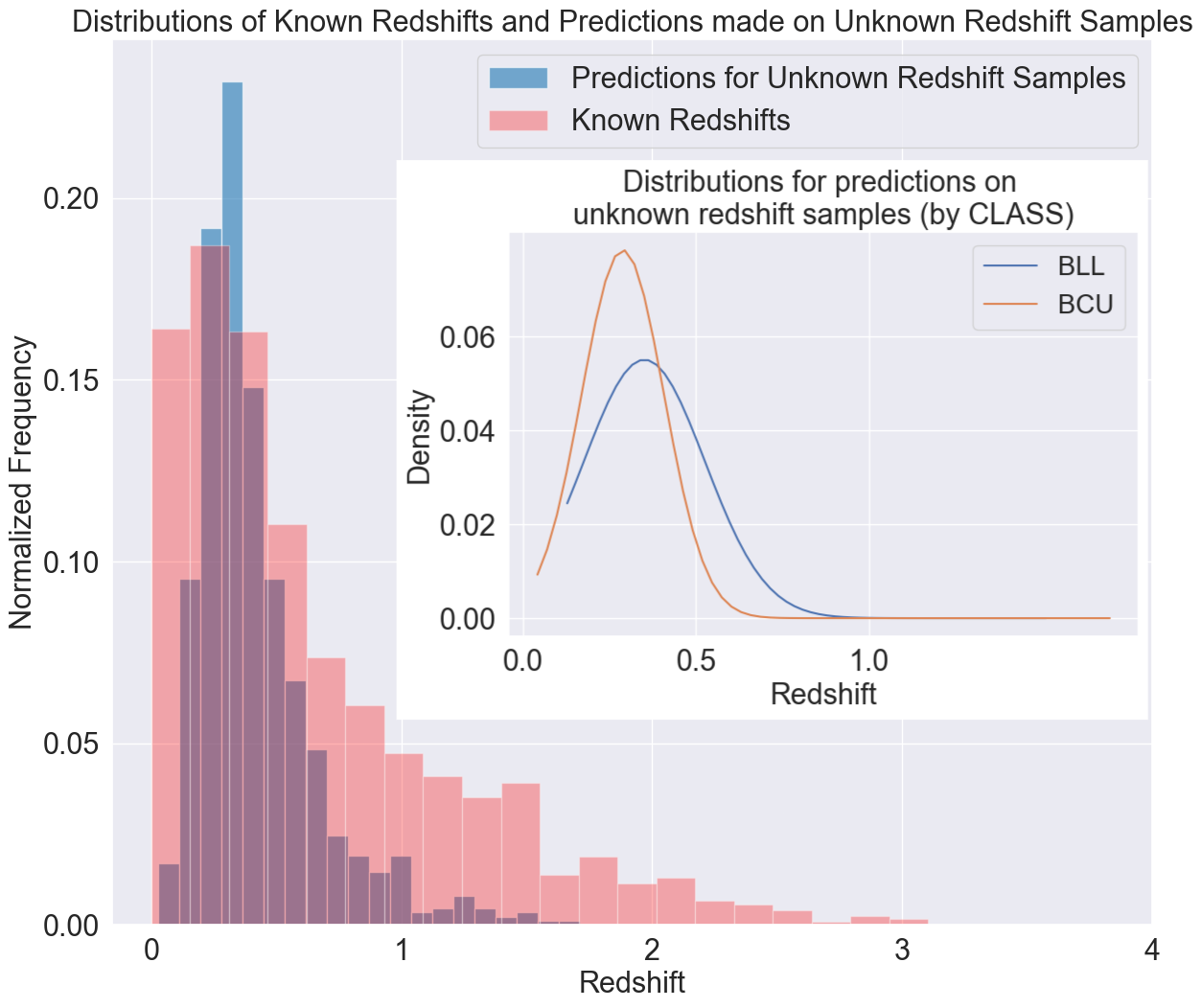}
    \caption{Variational Inference (Flipout Estimator) - Comparison between Known Redshift Samples and Predictions on Unknown Redshift samples using Histograms. The distribution of the redshift values for the predictions made on the unknown redshift samples, disaggregated by the "CLASS" feature. Only classes with more than 50 samples are represented.}
    \label{fig:Flipout_redshift1}
\end{figure*}

\begin{figure*}
    \centering
    \includegraphics[scale=0.325]{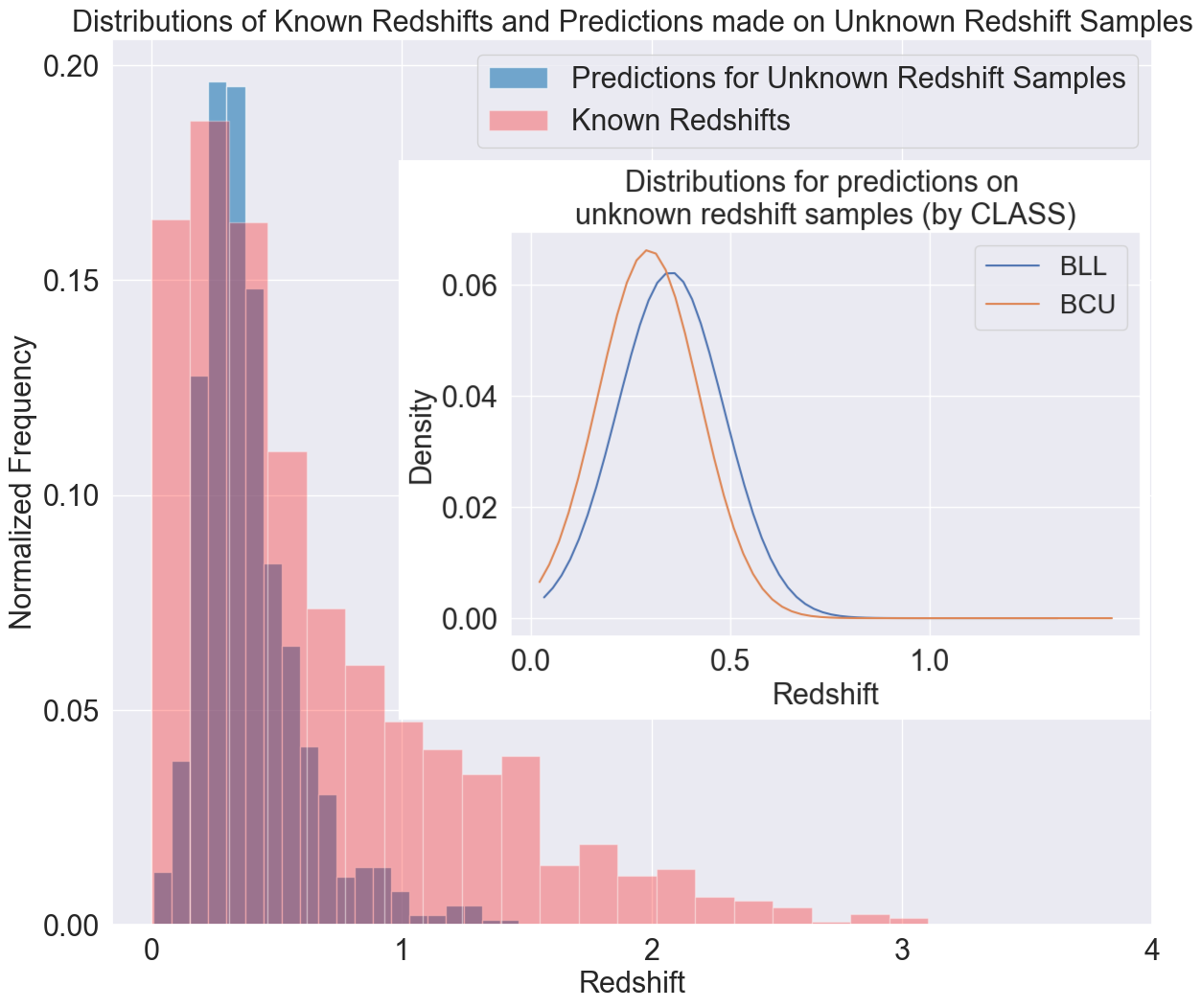}
    \caption{Variational Inference (Reparameterized Estimator) - Comparison between Known Redshift Samples and Predictions on Unknown Redshift samples using Histograms. The distribution of the redshift values for the predictions made on the unknown redshift samples, disaggregated by the "CLASS" feature. Only classes with more than 50 samples are represented.}
    \label{fig:Reparameterized_redshift1}
\end{figure*}

\begin{figure*}
    \centering
    \includegraphics[scale=0.325]{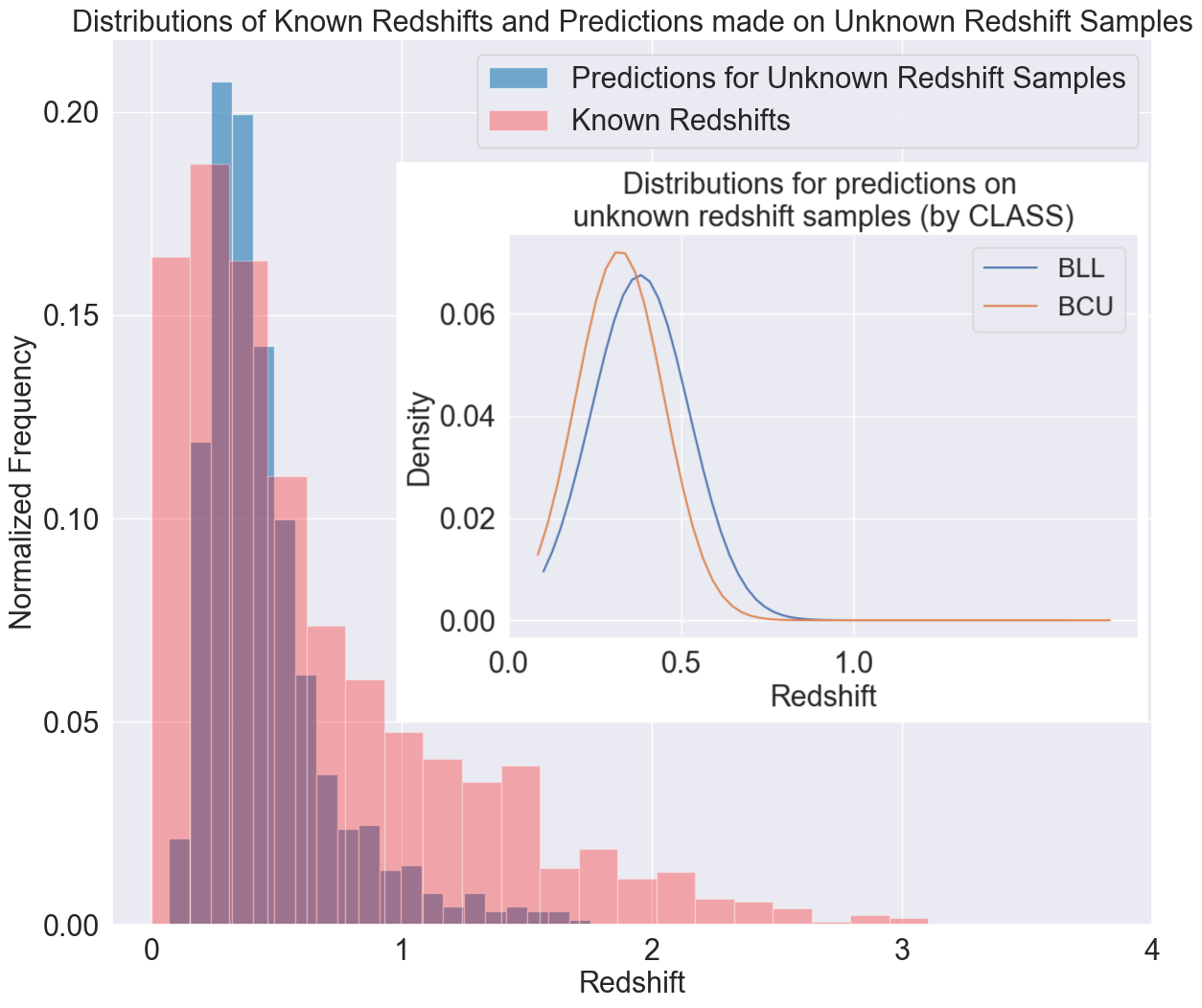}
    \caption{Frequentist model - Comparison between Known Redshift Samples and Predictions on Unknown Redshift samples using Histograms. The distribution of the redshift values for the predictions made on the unknown redshift samples, disaggregated by the "CLASS" feature. Only classes with more than 50 samples are represented.}
    \label{fig:Frequentist_redshift1}
\end{figure*}
\begin{figure*}
    \centering
    \begin{minipage}[b]{0.48\linewidth}
        \centering
        \subfloat[$z = 0.1860$]{\includegraphics[width=\linewidth]{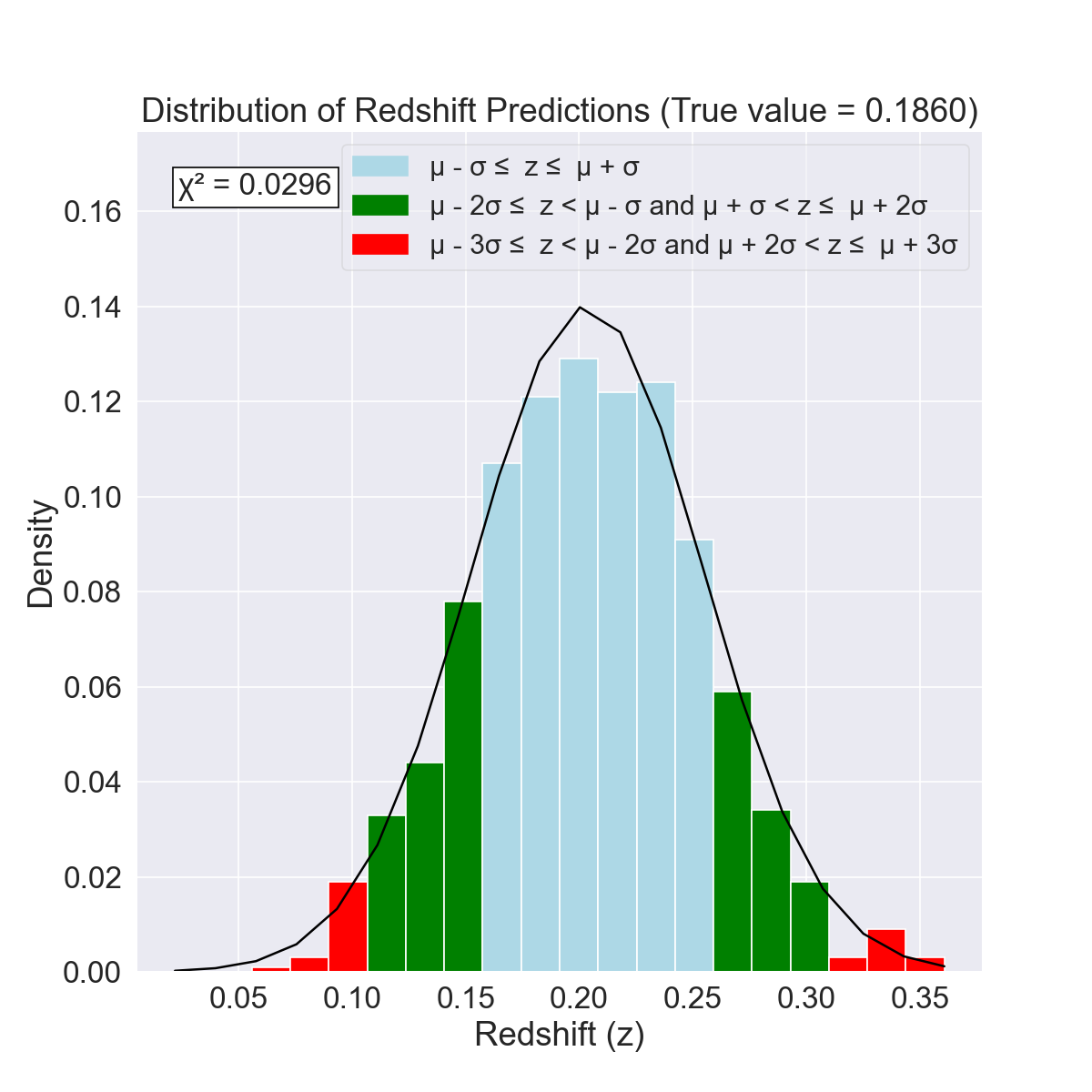}\label{fig:vi_0.1860}}
    \end{minipage}
    \hfill
    \begin{minipage}[b]{0.48\linewidth}
        \centering
        \subfloat[$z = 0.2973$]{\includegraphics[width=\linewidth]{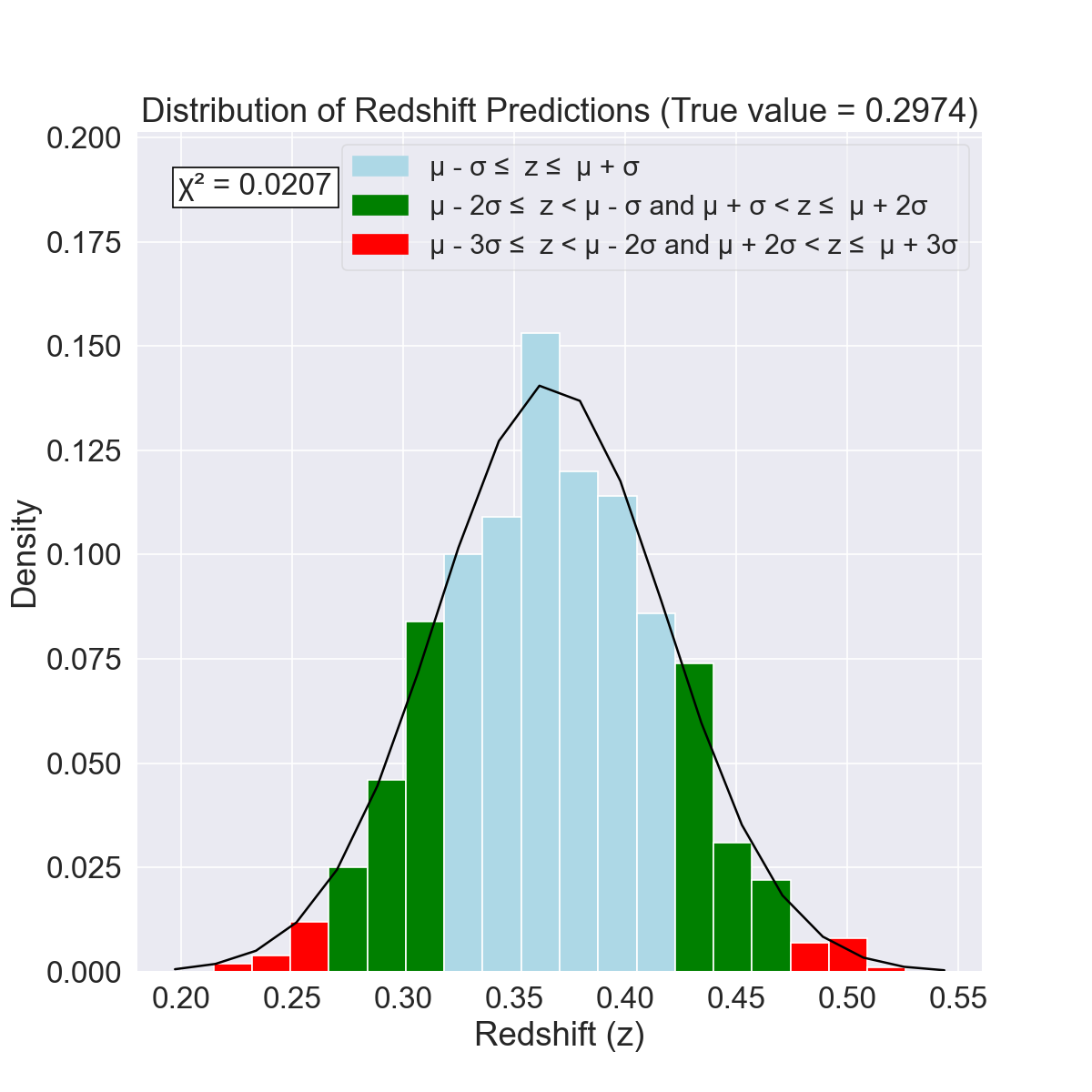}\label{fig:vi_0.2973}}
    \end{minipage}
    \begin{minipage}[b]{0.48\linewidth}
        \centering
        \subfloat[$z = 0.4469$]{\includegraphics[width=\linewidth]{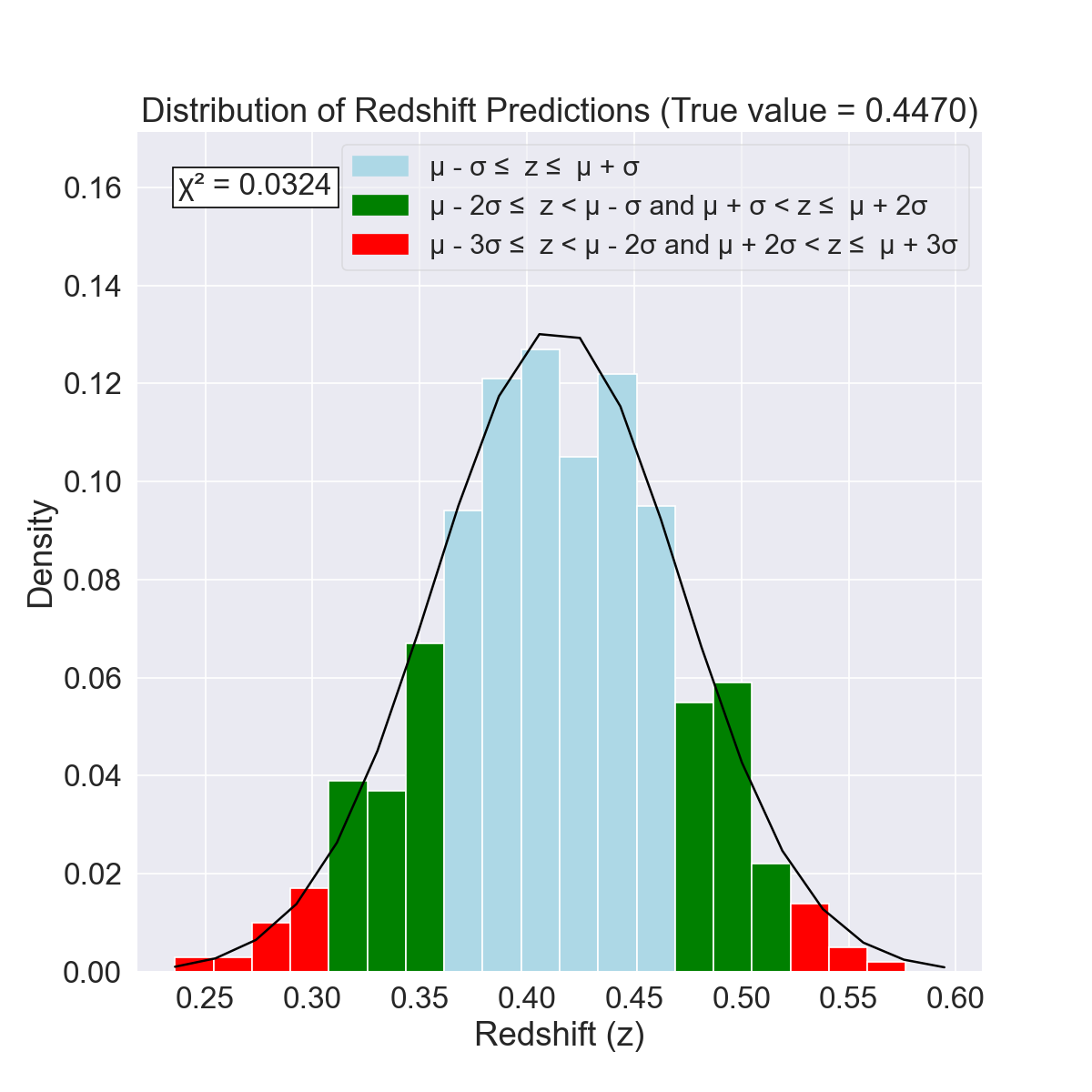}\label{fig:vi_0.4469}}
    \end{minipage}
    \hfill
    \begin{minipage}[b]{0.48\linewidth}
        \centering
        \subfloat[$z = 1.0140$]{\includegraphics[width=\linewidth]{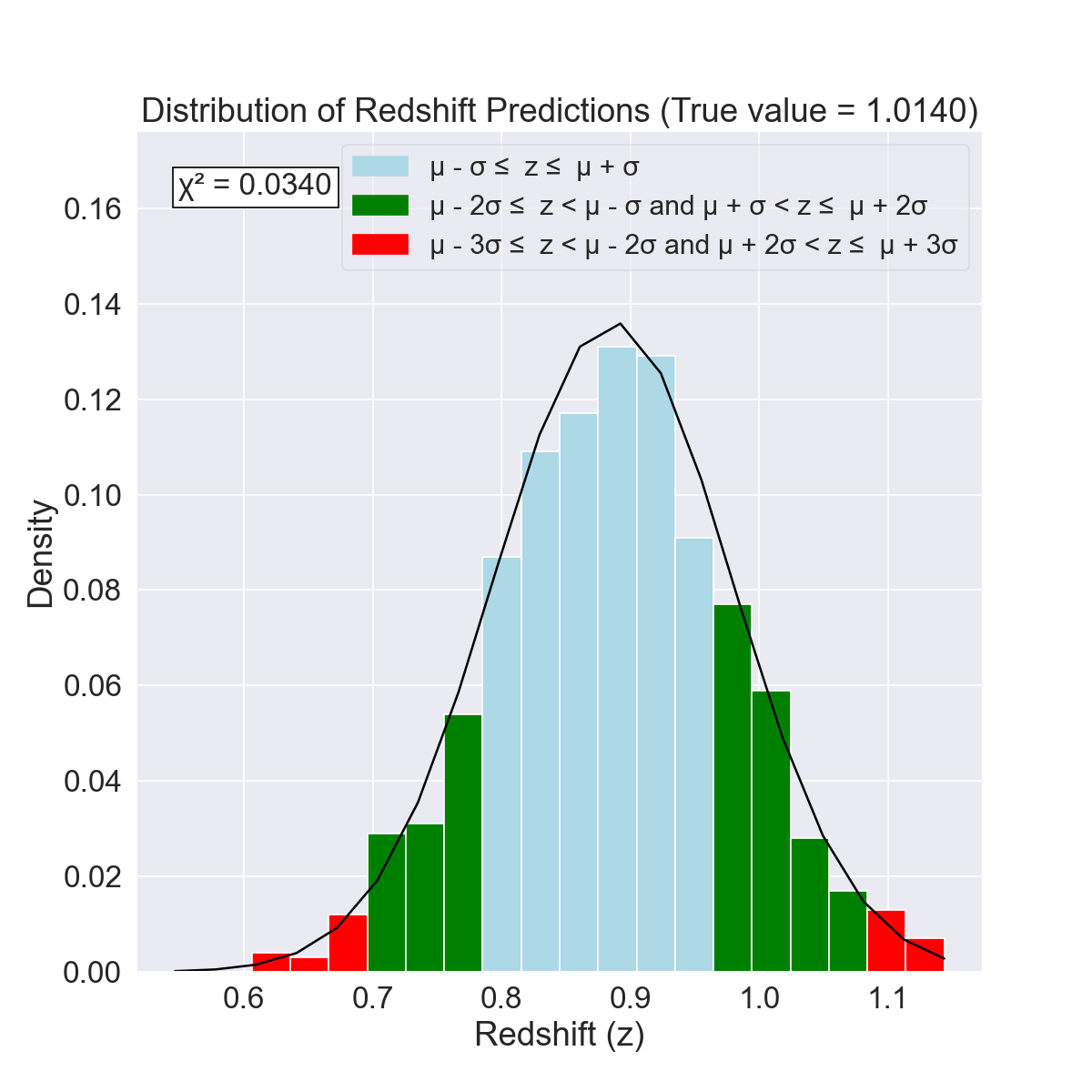}\label{fig:vi_1.0140}}
    \end{minipage}
    \caption{Evaluation of the uncertainty for variational inference using the flipout estimator. Redshift samples were evaluated 1000 times, and the resulting distribution for some of the known values is shown here. The distribution was fitted with a Gaussian PDF, and the values within 1$\sigma$, 1$\sigma$-2$\sigma$, and 2$\sigma$-3$\sigma$ from the mean were color-coded as cyan, green, and red, respectively.}
    \label{fig:vi_left_column_distributions}
\end{figure*}

\begin{figure*} 
    \centering
    \begin{minipage}[b]{0.48\linewidth}
        \centering
        \subfloat[$z = 0.1860$]{\includegraphics[width=\linewidth]{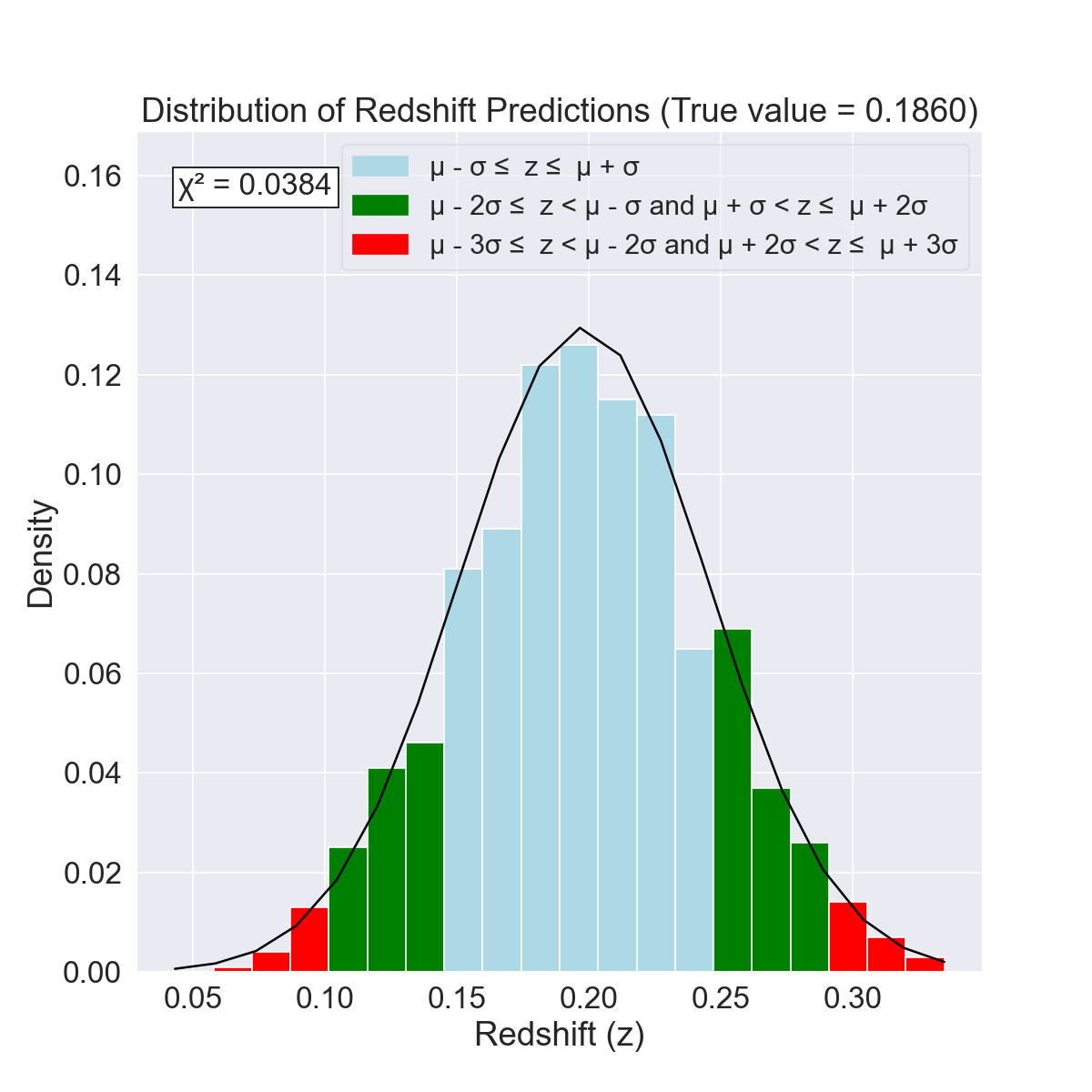}\label{fig:mc_0.1860}}
    \end{minipage}
    \hfill
    \begin{minipage}[b]{0.48\linewidth}
        \centering
        \subfloat[$z = 0.2973$]{\includegraphics[width=\linewidth]{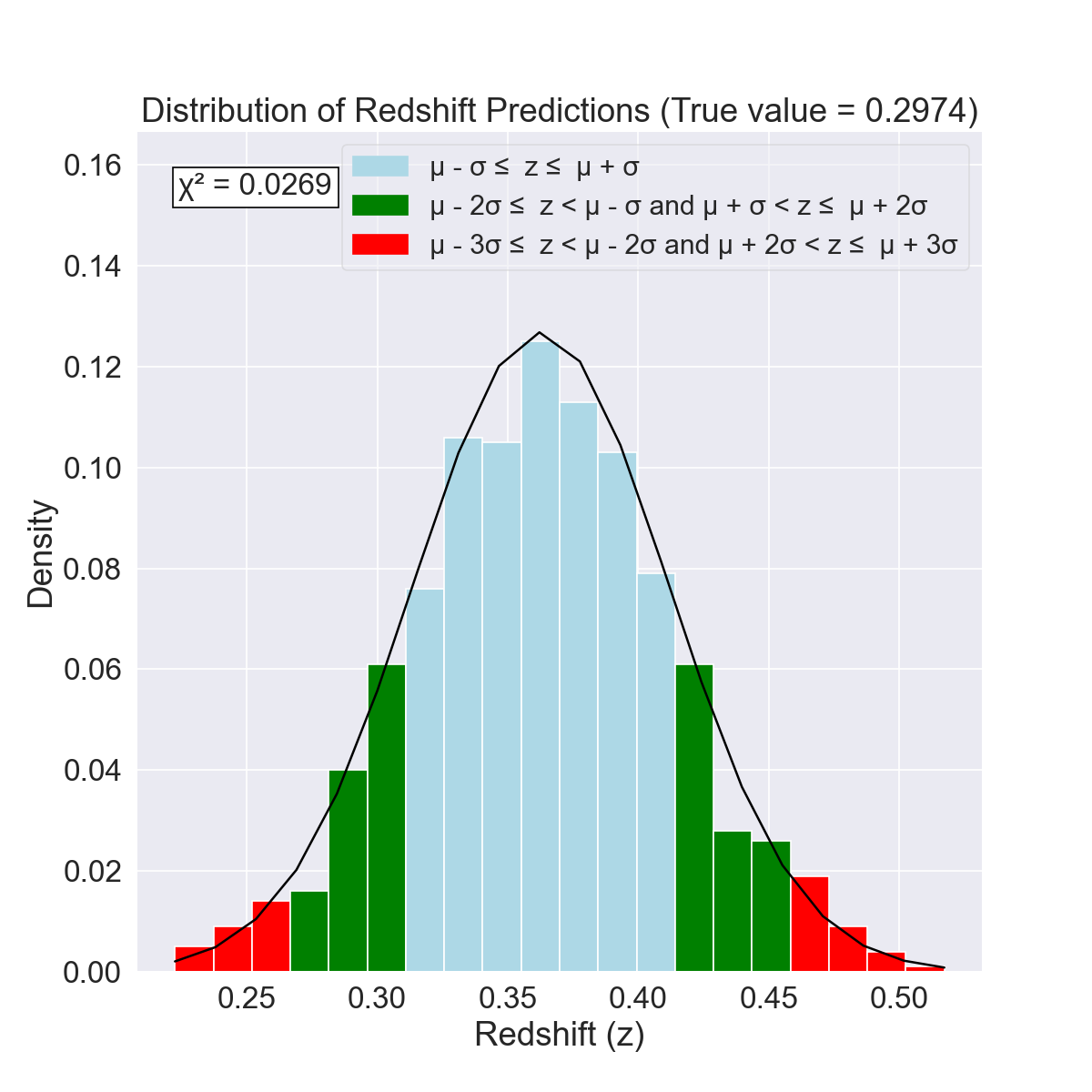}\label{fig:mc_0.2973}}
    \end{minipage}
    \vspace{0pt} 
    \begin{minipage}[b]{0.48\linewidth}
        \centering
        \subfloat[$z = 0.4469$]{\includegraphics[width=\linewidth]{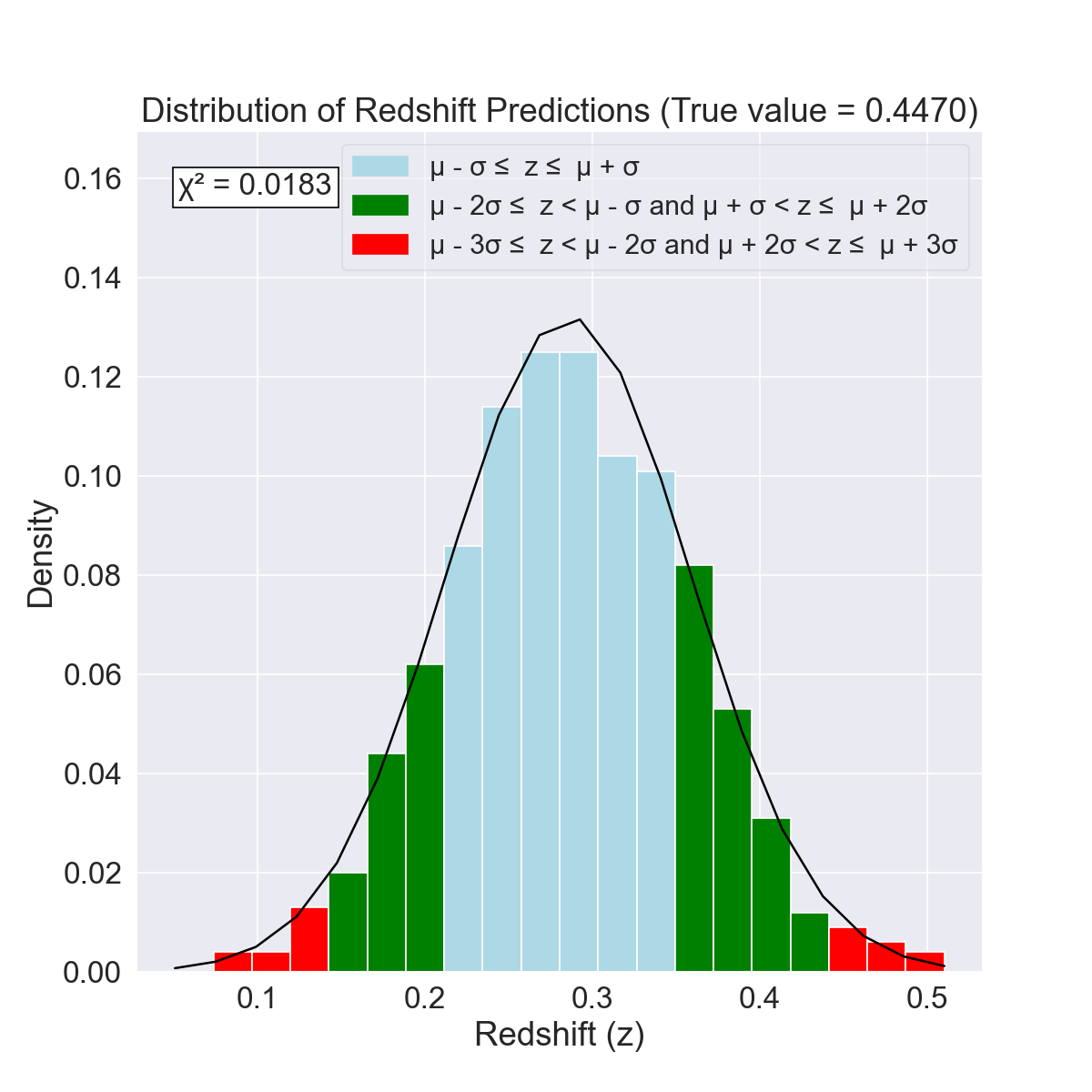}\label{fig:mc_0.4469}}
    \end{minipage}
    \hfill
    \begin{minipage}[b]{0.48\linewidth}
        \centering
        \subfloat[$z = 1.0140$]{\includegraphics[width=\linewidth]{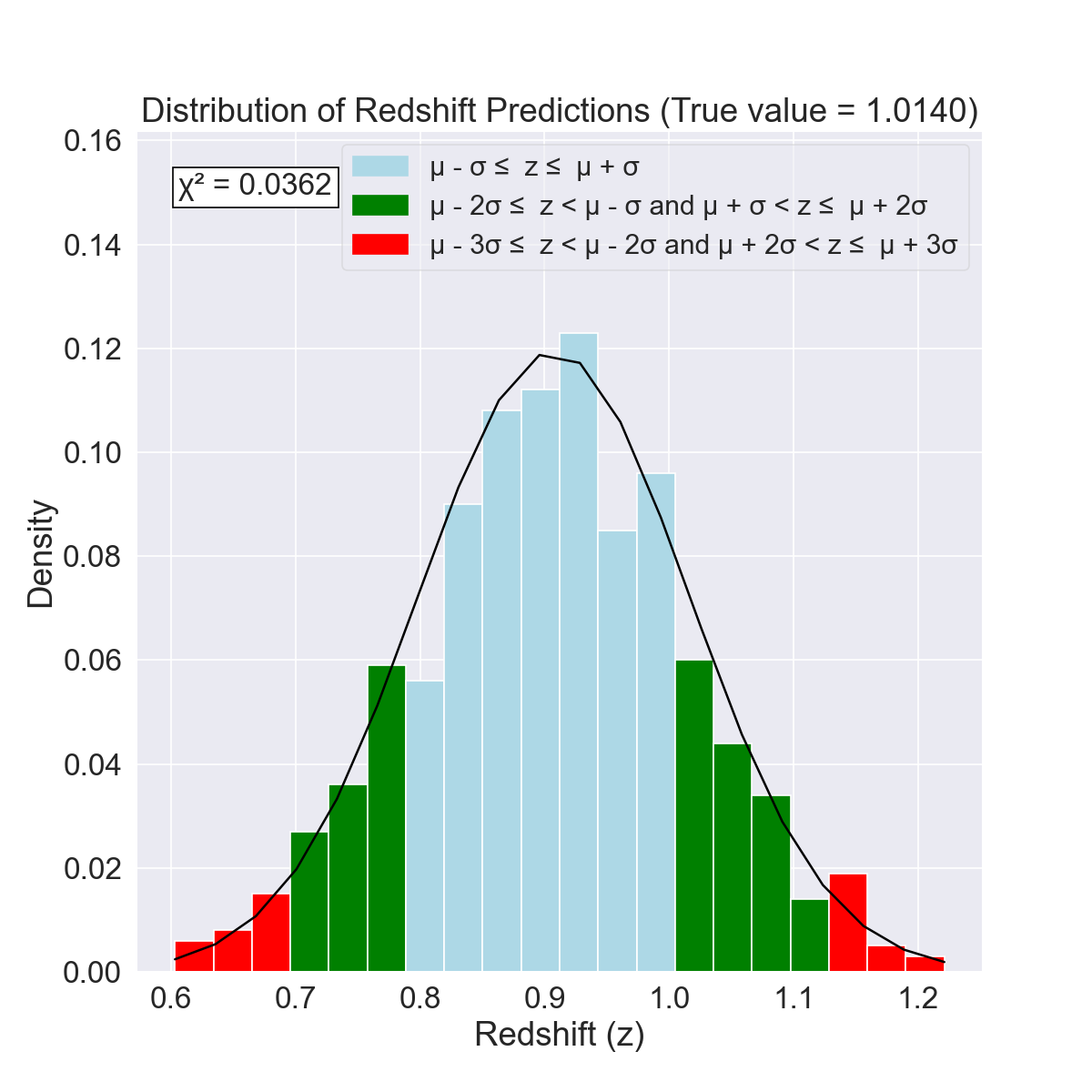}\label{fig:mc_1.0140}}
    \end{minipage}
    \caption{Evaluation of the uncertainty for variational inference using the reparameterization estimator. Redshift samples were evaluated 1000 times, and the resulting distribution for some of the known values is shown here. The distribution was fitted with a Gaussian PDF, and the values within 1$\sigma$, 1$\sigma$-2$\sigma$, and 2$\sigma$-3$\sigma$ from the mean were color-coded as cyan, green, and red, respectively.}
    \label{fig:vi_right_column_distributions}
\end{figure*}

\begin{table*}
\caption{Classwise distribution of the data considered for this study}
\centering
\label{tab:classwise}
\begin{tabular}{|c|c|c|c|}
\hline
& Known Redshift & Unknown Redshift & Total\\ \hline
BLL & 738 & 433 & 1171\\ \hline
BCU & 59 & 459 & 518\\ \hline
FSRQ & 390 & 0 & 390 \\ \hline
RDG & 26 & 4 & 30 \\ \hline
NLSY1 & 5 & 0 & 5 \\ \hline
AGN & 3 & 0 & 3 \\ \hline
CSS & 3 & 0 & 3 \\ \hline
Total & 1224 & 896 & 2120 \\ \hline
\end{tabular}
\end{table*}

\begin{table*}
    \centering
    \caption{Neural Network Architectures: Dropout of 0.25 between the hidden and output layers is common for each model}
    \label{tab:model_architecture_summary}
    \begin{adjustbox}{max width=\textwidth, center, scale=1}
    \begin{tabular}{lccc}
        \toprule
        Model & Hidden Layer & Output Layer & Estimator \\
        \midrule
        Frequentist & Dense (64 neurons) & Dense (1 neuron) & - \\
        \addlinespace
        Variational Inference & Dense (64 neurons) & DenseFlipout (1 neuron) & Flipout \\
        \addlinespace
        Variational Inference & Dense (64 neurons) & DenseReparameterization (1 neuron) & Reparameterization \\
        \bottomrule
    \end{tabular}
    \end{adjustbox}
\end{table*}

\begin{table*}
\caption{Performance Analysis: RMSE and Correlation Coefficient Comparison between Previous Approaches and Our Proposed Model. We make use of the mean predictions for the bayesian models to calculate the required metric}
\centering
\label{tab:performanceanalysis}
\begin{adjustbox}{width=\textwidth,center, scale=1}
\begin{tabular}{|c|c|c|c|c|c|c|}
\hline
& \cite{dainotti2021predicting} & \cite{narendra2022predicting} & \cite{coronado2023redshift} & Frequentist Algorithm & Variational Inference (Flipout)  & Variational Inference (Reparameterization) \\ \hline
RMSE & 0.432-0.438 & 0.458 & 0.46 & 0.415 & 0.406 & 0.438 \\ \hline
Correlation Coefficient & 0.704-0.718 & 0.74 & 0.71 & 0.784 & 0.777 & 0.778 \\ \hline
\end{tabular}
\end{adjustbox}
\end{table*}

\begin{table*}
  \centering
  \caption{Assessing redshift predictions using variational inference: Summary of True Values, Estimators, Confidence Intervals, and Variance for a random set of samples}
  \label{tab:redshiftprediction}
  \begin{adjustbox}{max width=18cm, scale=1.15}
    \begin{tabular}{lccccc}
      \toprule
      True Value & Estimator & 68.2\% CI & 95.4\% CI & 99.7\% CI & Variance \\
      \midrule
      0.1860 & Reparameterized & 0.1498--0.245 & 0.1022--0.2926 & 0.0546--0.3402 & 0.002 \\
      & Flipout & 0.1539--0.2533 & 0.1041--0.3031 & 0.0543--0.3529 & 0.002 \\
      \midrule
      0.2974 & Reparameterized & 0.3138--0.4124 & 0.2645--0.4617 & 0.2152--0.511 & 0.002 \\
      & Flipout & 0.3173--0.4177 & 0.2671--0.4679 & 0.2169--0.5181 & 0.002 \\
      \midrule
      0.4470 & Reparameterized & 0.2148--0.3588 & 0.1428--0.4308 & 0.0708--0.5028 & 0.005 \\
      & Flipout & 0.3563--0.4699 & 0.2955--0.5267 & 0.2427--0.5835 & 0.003 \\
      \midrule
      1.014 & Reparameterized & 0.7964--1.0154 & 0.6869--1.1249 & 0.5774--1.2344 & 0.011 \\
      & Flipout & 0.7924--0.979 & 0.6991--1.0723 & 0.6058--1.1656 & 0.008 \\
      \bottomrule
    \end{tabular}
  \end{adjustbox}
\end{table*}

\begin{table*}
  \caption{Redshift Prediction Summary Statistics}
  \label{tab:summary_statistics}
  \centering
  \begin{adjustbox}{max width=\textwidth, center, scale=1}
  \begin{tabular}{lccccccc}
    \toprule
    \multirow{2}{*}{Method} & \multicolumn{3}{c}{Known Redshift samples} & \multicolumn{3}{c}{Unknown Redshift samples} \\
    \cmidrule(lr){2-4} \cmidrule(lr){5-7}
    & Mean Prediction & Range & $\sigma$ & Mean Prediction & Range & $\sigma$ \\
    \midrule
    Frequentist Model & 0.559 & 0.04 - 1.99 & 0.372 & 0.455 & 0.07 - 1.77 & 0.258 \\
    Variational Inference (Flipout Estimator) & 0.581 & 0.027 - 2.11 & 0.382 & 0.415 & 0.0251 - 1.71 & 0.246 \\
    Variational Inference (Reparameterization Estimator) & 0.526 & 0.0004 - 1.82 & 0.332 & 0.393 & 0.0089 - 1.47 & 0.207 \\
    \bottomrule
  \end{tabular}
  \end{adjustbox}
\end{table*}

\section{Introduction}
Redshift, denoted as "z", is a measure of the displacement of spectral lines towards longer wavelengths in the electromagnetic spectrum. This phenomenon arises due to the expansion of the universe, stretching the wavelength of light emitted by distant celestial objects. Redshift estimation plays a fundamental role in understanding the properties of these objects, including their distance, cosmological evolution, and the nature of the universe itself. In the realm of astrophysics, redshift estimation traditionally relies on spectroscopic measurements, where the light emitted by celestial objects is dispersed into its constituent wavelengths, revealing characteristic absorption or emission features. However, spectroscopic observations are often constrained by limited observational time, expensive resources, and the technical limitations of spectrographs. Consequently, obtaining spectroscopic redshift measurements for a large number of objects, as required by comprehensive surveys, becomes challenging and impractical. \\\\
The Fermi Gamma-ray Space Telescope (Fermi-LAT) has revolutionized the study of high-energy gamma-ray sources and contributed significantly to our understanding of the universe. The Fermi-LAT observatory observes celestial objects in gamma-ray wavelengths. However, efficiently extracting redshift information solely from gamma-ray observations poses a challenge 
as these observations are devoid of any spectral line, besides that of the 511 keV feature \cite{skinner2010positrons}. Therefore, the sole viable approach to gauge the distance involves linking the gamma-ray emitter with a recognized source that exhibits absorption or emission lines in other wavelengths, thereby enabling the calculation of redshift. 
The majority of discrete sources detected by Fermi/LAT are blazars, which consist of flat-spectrum radio quasars (FSRQs) exhibiting distinct optical emission lines over a broad-band continuum, and BL Lacs (BLLs), characterized by weak or absent emission line signatures \citep[see][and references therein]{2020ApJ...891..120B}. This indicates that while it may be relatively easier to estimate the redshifts of FSRQs, the redshift evaluation for BL Lacs is a complex and often computationally expensive task as it necessitates extensive optical spectroscopic observations along with comprehensive multi-wavelength observations involving diverse astronomical facilities.\\\\
To address these challenges, astronomers have turned to ML and DL techniques \cite{dainotti2021predicting, narendra2022predicting, coronado2023redshift}, which have demonstrated remarkable success. The study done by \cite{dainotti2021predicting} is one of the initial works in estimating the redshift of $\gamma$-Ray loud AGNs. The authors make use of an ensemble-based approach that combines standard regression algorithms such as Random Forest, XG Boost, Big LASSO, and Bayes GLM to estimate the redshift of the corresponding input target. The authors make use of a $10$ cross-fold validation technique iterated over $10$ times to report a correlation coefficient (r) ranging from $0.704$ to $0.718$. Moreover, they also reported a root-mean-squared error (RMSE) ranging from $0.432$ to $0.438$.\\\\
\cite{narendra2022predicting} is an advancement of \cite{dainotti2021predicting}. The authors employed a similar ensemble-based technique as observed in \cite{dainotti2021predicting}, however, the only difference besides an increase in the data points and the feature vector is the choice of machine learning models. The authors report an RMSE value of $0.212$ when the sample size is $111$ and $0.458$ when the sample size is $1112$. As RMSE is inversely proportional to the number of samples used during evaluation, it can not be considered the best evaluation metric to compare different algorithms unless the sample size is the same across the algorithms. Also, the authors report a correlation coefficient of r $\approx 0.74$ in both of the aforementioned cases \\\\
In \cite{coronado2023redshift}, the author makes use of the 4LAC DR3 catalog, which is an updated version of the data used in \cite{dainotti2021predicting} and \cite{narendra2022predicting} with multiple additional features and a significant increase in the number of data points. To optimally use both the numerical as well as categorical features, the author relies on the CatBoost algorithm which is a boosted decision tree-based algorithm capable of dealing with the categorical data. The author employs a $5$-cross validation technique to make effective use of the limited data. The reported "RMSE" and "r" values in this study are $0.46$ and $0.71$ respectively. Similar to \cite{dainotti2021predicting} and \cite{narendra2022predicting}, the author also experiments with an ensembled approach having combined eight different algorithms, however, the performance of the CatBoost model is reported to be significantly better than what was observed in the ensembled algorithm.\\\\
Considering the limited number of studies conducted on this topic, none of which account for the uncertainty of the predicted redshifts, in this manuscript, we introduce an algorithm that employs a multi-layer perceptron with a single hidden layer as the foundational model which when modified using variational inference allows us to not only quantify uncertainty but also augment our results.



\section{Methodology}


\subsection{Data Collection and Processing}
Since its launch in 2008, the Fermi Gamma-Ray Space Telescope's onboard instrument called the LAT has been continuously monitoring the high-energy sky \citep{2009ApJ...697.1071A}. In this study, we utilize the Fermi fourth catalog of active galactic nuclei (AGNs) data release 3 (4LAC-DR3; \citet{ajello2022fourth, 2022ApJS..263...24A}).
The catalog comprises $3407$ individual sources, of which $1806$ sources have known redshifts. Each source is characterized by a set of $41$ different features with randomly missing values reported in this catalog. Following \cite{coronado2023redshift}, we shortlist a set of $24$ features for our study. Some of the features such as "SED\_class", "Highest\_energy" and "Unc\_LP\_beta" have a number of missing values. After sufficient experimentation with different imputing techniques, feature removal, and data removal we proceed with the removal of the data points with missing values for the "Highest\_energy" and "Unc\_LP\_beta" features. On the other hand, the missing values for the "SED\_class" are imputed using the mode estimation technique or, most frequent categorical value imputation \cite{lin2020missing}. To carry out the imputation process, we make use of sklearn's \cite{pedregosa2011scikit} SimpleImputer  with appropriate arguments like setting strategy to most\_frequent. This leaves us with $1224$ data points (For detailed data distribution refer to Table \ref{tab:classwise}) for our study with $90\%$ of the data used for training and $10\%$ of data used for validation and testing purposes equally divided among each other. It is crucial to note that the concept of the \emph{validation data split} refers to the division of data used for evaluating and refining a deep learning model during its training process. This division serves as a means to optimize the model's performance and make necessary adjustments. By subjecting the trained model to the validation set, we gain valuable insights into its ability to generalize on unseen data. The model's performance on the validation set can be regarded as a reliable indicator of its performance on entirely new data at each training epoch. This evaluation helps in identifying potential issues, such as overfitting, which can significantly affect the model's effectiveness in real-world applications. It allows us to make unbiased estimations of critical hyperparameters, such as the number of neurons in the hidden layer or the dropout rate, essential for optimizing the model's performance. The collected data consists of a number of numerical and categorical features. To deal with categorical data, we convert them to an integer-valued array using Sklearn's ordinal encoder. Next, all the numeric data is normalized using the StandardScaler provided by Sklearn \cite{scikit-learn, sklearn_api}. Prior to standard scaling, all numeric features except "Frac\_Variability", "GLAT", "GLON", "LP\_Index", "LP\_beta", "PL\_Index", "Unc\_Flux1000" and "Unc\_PL\_Index" undergo log transformations. After pre-processing we are left with a total of $1224$ samples with  known redshift values and $896$ samples with unknown redshifts. Please refer Table \ref{tab:classwise} for class-wise distribution of the samples. The feature engineering and data engineering and proposed algorithms are implemented using python 3.7. The pandas library \cite{mckinney-proc-scipy-2010} is used to read the dataframes from the files and store them and once the input features are identified, we store them using numpy \cite{harris2020array} arrays, in order to feed them into our TensorFlow \cite{tensorflow2015-whitepaper} models.

\subsection{Model Architecture and Uncertainty Quantification}

 

In this study, we propose a multi-layer perception \cite{murtagh1991multilayer, noriega2005multilayer, baum1988capabilities} with a single hidden layer having $64$ neurons. A multi-layer perceptron, often abbreviated as MLP, is a feed forward Neural Network with atleast 3 layers including the standard input, hidden and output layer. Every layer has multiple nodes/neurons in it, which along with the number of hidden layers define the complexity of the model. Though there are many standard techniques to define the number of neurons in every hidden layer, in this study due to the simplicity of our model, we come up with the value of $64$ after sufficient experimentation. These MLPs are fully connected,  implying that every node in layer "i" connects to each node in the subsequent layer "j" through a weight value denoted as $w_{ij}$. The learning process is facilitated by adjusting the values of these weights as the data is processed, guided by the error between the MLP output and the target value. Further, to avoid overfitting, we introduce a dropout \cite{srivastava2014dropout, cai2019effective, srinivas2016generalized} of $0.25$ in the hidden layer. This ensures that, during training, at any point in time, a neuron will be inactive with a probability of $0.25$. This prevents the network from relying too heavily on specific neurons and encourages more robust, generalized learning.
Next, to ensure non-linearity within the model we apply ReLU - a widely used activation function \cite{fukushima1975cognitron}. In the output layer, we utilise the softplus activation function \cite{DUBEY202292} which is just a smooth continuous version of ReLU. Placing an activation function at the end of each layer ensures that the layer's output undergoes a non-linear transformation before being passed to the next layer. This is crucial in enabling the algorithm to learn and capture non-linear dependencies between the input and the output. For the loss function, we employ the "Mean Absolute Error" (MAE) \cite{hodson2022root}. This baseline model treats its parameters as point estimates and hence we refer to it as the "frequentist" model. Moreover, to account for uncertainty, we employ the method of variational inference to modify our frequestist model using two different estimators, as discussed below. A summary of the architectures for the three models is listed in Table \ref{tab:model_architecture_summary} \\\\
Variational inference is a technique that aims to approximate the true but often intractable posterior distribution of the model's parameters (weights and biases) given the observed data \cite{shridhar2019comprehensive, jospin2022hands}. Instead of directly calculating the posterior, which is either challenging or impossible in complex models, variational inference introduces an approximating distribution (usually a known and tractable distribution).\\\\
To achieve this, a prior distribution is assigned to the model's parameters, representing our initial beliefs about their values. As data is observed, the prior is updated using Bayes' rule to obtain the posterior distribution. However, directly calculating the posterior is intractable for many models, especially neural networks. Thus, an optimization problem is formulated: we seek the closest approximating posterior distribution (in terms of the Kullback-Leibler (KL) divergence) that can be efficiently computed \cite{10.5555/1162264}. Both the prior and the approximating distributions are chosen as the Normal distribution, due to its desirable properties, like being a conjugate prior to itself.\\\\
Unlike traditional neural networks that rely on point estimates, variational inference provides a more meaningful measure of uncertainty and captures the complexity of the posterior distribution through this probabilistic approach. We use TensorFlow Probability \cite{DBLP:journals/corr/abs-1711-10604} and Keras \cite{chollet2015keras} to implement the proposed models. There are multiple methods to implement variational inference using tensorflow probability, however, we proceed with the DenseFlipout and DenseReparamterization layers. In both of these methods, the layers implement the Bayesian variational inference counterpart to a Dense layer by drawing the parameter values from distributions. An important difference between both these layers is that the flipout estimator uses roughly twice as many floating point operations as the reparameterization estimator. (Refer \cite{wen2018flipout} and \cite{kingma2013auto} for more information on both of these layers).\\\\
To quantify uncertainty, each sample is evaluated $1000$ times and the uncertainty is captured using the variance of the predictions. The resulting mean from the $1000$ iterations is considered as the prediction of the Bayesian model. Due to the Bayesian nature of the variational inference algorithms, the output prediction at every iteration is an independent and identically distributed Gaussian sample. Having set the output predictions to be normally distributed for a fixed data point, we then calculate the mean and standard deviation of the predictions for each sample. As evident from the theory of Gaussian distributions, we then make use of the standard $3$ sigma rule to come up with a possible range of redshifts containing the true value of the redshift with an associated confidence level. Although this rule comments on the confidence levels being, $68.2, 95.4$, and $99.7$ percent for 1, 2 and 3 standard deviations from the mean, respectively, it is easy to generalize it for any range of values depending upon the allowed tolerance.



\subsection{Training and Validation}

Considering the computational requirement to train the algorithm, we make use of Google Collaboratory, a cloud-based jupyter environment for model training. An important aspect of any Machine or Deep Learning algorithm is its reproducibility. To ensure this, we train our algorithms on a fixed random seed over a maximum of $2500$ epochs, and include the data splits pertaining to the training, validation and testing sets in our GitHub repository. To reduce the computational overhead and avoid overfitting, we introduce early stopping \cite{caruana2001overfitting} with a validation patience of $100$, and as a result, the proposed variational inference models stop after $1170^\text{th}$ and $390^\text{th}$ epoch respectively as shown in Figure \ref{fig:flipout} and \ref{fig:reparameterization} respectively.\\\\
As evident from Figure \ref{fig:reparameterization}, during the initial $100$ epochs, the rate of decrease in "loss" and "RMSE" for both the training and validation data points is high. However, at later stages, it tends to saturate. This indicates that there's a very high probability of having no further decrease in the loss. Having said this, the use of early stopping ensures that the algorithm stops its training once the rate of decrease in the validation loss tends to zero. This helps in avoiding unnecessary computations. Also, in Figure \ref{fig:flipout}, we observe that at later stages there's a decrease in training loss, on the other hand, the validation loss tends to saturate and even increases in further epochs. This behavior results in overfitting of the algorithm, if not stopped at the correct time, and the introduction of early stopping ensures the same.\\\\
To optimize the algorithm, we make use of "Adam" \cite{kingma2014adam} which is one of the widely used optimizers in the Deep Learning community with a learning rate of $10^{-3}$. One of the primary reasons for its popularity is that it incorporates momentum (for which we use the default values defined in TensorFlow) and is a variant of the AdaGrad optimizer, which facilitates quicker convergence.

\section{Results and Discussion}
Blazars emitting $\gamma$-rays with known redshifts significantly contribute to our understanding of several fundamental aspects of cosmic phenomena. Determining their redshifts aids in constraining the nature of the Extragalactic Background Light \citep[e.g.,][]{2019MNRAS.486.4233A, 2013APh....43..112D, 2012Sci...338.1190A}. Additionally, these blazars shed light on the structures of intergalactic magnetic fields \citep[e.g.,][]{2023ApJ...950L..16A, 2015ApJ...814...20F, 2010MNRAS.406L..70T} and the universe's star formation history \citep{2018Sci...362.1031F, 2016MNRAS.463.1068R, 2012ApJ...755..164A}. 
Also, by computing the luminosity function, we can estimate the evolution of blazars over cosmic time \citep{1995ApJ...452..156C, 2012ApJ...751..108A}. This, in turn, can lead to the constraining of fundamental cosmological parameters \citep{2019ApJ...885..137D, 2019ApJ...882...87Z}.\\\\
The study contributes by providing an algorithm that rigorously estimates the possible range of redshifts with an associated confidence. To assess the effectiveness of our model, we conducted evaluations using entirely new and unseen data, referred to as the test data. Since our study focuses on a regression problem, we utilized the "Root Mean Squared Error" (RMSE) as one of our evaluation metrics. The Root Mean Squared Error (RMSE) calculates the square root of the average of the squared differences between predicted values and actual values in the test data. A lower RMSE generally indicates better model performance, as it signifies smaller prediction errors. However, the RMSE value is influenced by the number of samples. Therefore, we also utilized the "correlation coefficient" to evaluate our model. The correlation coefficient measures the strength and direction of the linear relationship between two variables. A higher correlation coefficient indicates a better alignment between the predicted and actual values, demonstrating the model's ability to capture the underlying patterns in the data. Table \ref{tab:performanceanalysis} clearly shows that our proposed algorithm yields improved results when compared to existing studies, with a maximal increase in the correlation coefficient of around $0.07$.\\\\
Additionally, Table \ref{tab:redshiftprediction} presents a comparison between the actual redshift values and the predicted range of redshifts at fixed confidence levels for randomly selected data points from the test dataset. The table clearly demonstrates that in the majority of cases, the true redshift value falls within the interval associated with a confidence level of $95.4\%$. Although in table \ref{tab:redshiftprediction}, we focus on the specific confidence levels, the range can be easily calculated for different confidence levels based on a real multiple of the standard deviation.\\\\
Figures \ref{fig:flipout_scatter}, \ref{fig:reparameterized_scatter} and \ref{fig:frequentist_scatter} present scatter plots that showcase the relationship between predicted and true redshifts obtained from various models. While it is evident that the predicted redshifts tend to be slightly lower than the actual values in many instances (a trend also observed in \cite{coronado2023redshift}, albeit with more scattered points), the incorporation of uncertainty and confidence levels addresses this issue. By utilizing a 3-sigma interval of the mean with a confidence level of $99.7\%$, the majority of true values fall within this range - an analysis reveals that for all the samples with a known redshift, the true value falls within the $99.7\%$ confidence interval for $63\%$ of the samples using each method of variational inference. This enables astronomers to make informed decisions regarding the reliability of the algorithm's predictions, considering the desired confidence level and width interval at any given point. Figures \ref{fig:Flipout_redshift}, \ref{fig:Reparameterized_redshift} and \ref{fig:Frequentist_redshift} provide similar insights. Additionally, the figures highlight the algorithm's limitation in regressing lower redshifts. However, due to the associated uncertainty and the range of predictions provided by Variational Inference, the lower redshifts are accounted for within the predicted range. This aspect of our proposed algorithm ensures that the true value is captured with a sufficiently high probability, depending on the allowed confidence level.\\\\
As illustrated in Figures \ref{fig:Flipout_redshift1}, \ref{fig:Reparameterized_redshift1} and \ref{fig:Frequentist_redshift1}, the predictions made on the  samples with an unknown redshift by the frequentist model, the flipout estimator model and the reparamterization estimator model follow distributions similar to that of the predictions made on the known redshift samples with mean values of $0.455$, $0.415$ and $0.393$, standard deviations of $0.258$, $0.246$ and $0.207$ and redshift values ranging from $0.07$-$1.77$,  $0.0251$-$1.71$ and $0.0089$-$1.47$, respectively (Table \ref{tab:summary_statistics}).\\\\ 
Figures \ref{fig:vi_left_column_distributions} and \ref{fig:vi_right_column_distributions} display histograms corresponding to the data presented in Table \ref{tab:redshiftprediction}. As evident from the figures, the predicted set of values for every redshift correspond to a Gaussian distribution which confirms the inclination of the implemented algorithm with the theory and hence allows us to efficiently estimate the uncertainty associated with the range of predictions.\\\\
Also, as seen in Table \ref{tab:classwise} and Figures \ref{fig:Flipout_redshift1}, \ref{fig:Reparameterized_redshift1} and \ref{fig:Frequentist_redshift1}, the predicted redshift class is mostly composed of BL Lacs and BCUs. These results are plausible because BL Lacs are strong gamma-ray emitters with weak or no emission lines, which makes estimating their redshifts very difficult. Similarly, the BCUs are unclassified sources whose classification is challenging,  as optical spectra or MWL observations required for a robust classification are not available. However, several studies based on machine learning predict that the majority of these sources are likely to be BL Lacs \cite[see e. g.,][]{2023ApJ...946..109A,Kang_2019}. 

\section{Conclusion}
This study introduces a straightforward yet highly effective algorithm for redshift estimation using solely Gamma-Ray observations.The proposed algorithm shows improvements over existing methods, achieving  significantly low RMSE values of $0.415, 0.406$, and $0.438$ in its frequentist, variational inference (flipout), and variational inference (reparameterization) variants respectively. To further validate our results, we also employ the correlation coefficient as a complementary metric. Remarkably, we observe a substantial improvement in the correlation coefficient, with values increasing from $0.74$ to $0.784, 0.777$, and $0.778$ for the respective algorithms, thus demonstrating the advantage of our proposed method. In addition to robust redshift regression, our algorithm addresses the associated uncertainty by providing an estimated range of potential redshift values based on the desired confidence level. Notably, for highest confidence interval ($99.7\%$), the predictions of our algorithm encompass the true redshifts for the majority of the samples. This uncertainty quantification feature adds significant value to the algorithm's predictions and helps users to make informed decisions based on their desired confidence level. Furthermore, we extend the application of our algorithm to predict unknown redshifts in the 4LAC-DR3 catalog, utilizing variational inferences. This allows us to provide corresponding uncertainties alongside the predicted redshifts, enhancing the reliability and applicability of our algorithm in real-world scenarios. 

\section*{Acknowledgement}
We thank the anonymous referee for a careful and thorough review of this paper, which helped us improve the quality of the work.

\section*{Data Availability}
The data utilized in this paper can be accessed by the public through the Fermi Science Support Center (FSSC) of NASA's Goddard Space Flight Center. Furthermore, we have made both the code and the resulting data openly available on our \href{https://github.com/abhimanyu911/redshift-regression-with-uncertainty.git}{Github repository (https://github.com/abhimanyu911/redshift-regression-with-uncertainty.git)} for public access.



\bibliographystyle{mnras}
\bibliography{example} 








\bsp	
\label{lastpage}
\end{document}